\documentclass{svproc}
\usepackage{url}
\usepackage[utf8]{inputenc} 
\usepackage[T1]{fontenc}    
\usepackage{hyperref}       
\usepackage{url}            
\usepackage{booktabs}       
\usepackage{amsfonts}       
\usepackage{nicefrac}       
\usepackage{microtype}      
\usepackage{xcolor}         
\usepackage{graphicx}
\usepackage[ruled,vlined,linesnumbered]{algorithm2e}

\SetCommentSty{mycommfont}

\usepackage{tabulary}
\usepackage{booktabs}
\usepackage{multirow}
\usepackage{caption}
\usepackage{subcaption}
\usepackage{media9}

\hypersetup{
    colorlinks=true,
    linkcolor=blue,
    filecolor=red,      
    urlcolor=blue,
    citecolor=red,
    pdfpagemode=FullScreen,
    }

\begin{document}
\mainmatter              
\title{Can one hear the position of nodes?}
\titlerunning{Can one hear the position of nodes?}  
%
\author{Rami Puzis}
\authorrunning{Puzis} 

\institute{Department of Software and Information Systems Engineering,\\
            Ben-Gurion University of the Negev, Beer-Sheva, Israel \\
\email{puzis@bgu.ac.il}
}

\maketitle              

\begin{abstract}
Wave propagation through nodes and links of a network forms the basis of spectral graph theory. 
Nevertheless, the sound emitted by nodes within the resonating chamber formed by a network are not well studied. 
The sound emitted by vibrations of individual nodes reflects the structure of the overall network topology but also the location of the node within the network. 
In this article a sound recognition neural network is trained to infer centrality measures from the nodes' wave-forms.
In addition to advancing network representation learning, sounds emitted by nodes are plausible in most cases. 
Auralization of the network topology may open new directions in arts, competing with network visualization.  
\end{abstract}

\section{Introduction}
\label{sec:introduction}



Representation learning in graphs, a.k.a embedding, facilitates variety of downstream analysis tasks such as node classification~\cite{ma2019riwalk}, link prediction~\cite{mallick2019topo2vec}, community detection~\cite{sun2019vgraph}, network classification~\cite{riesen2009graph}, and more.
One challenging application of representation learning is inference of the importance (centrality) of nodes in a network according to their position in the network topology. 
There were many centrality measures defined over the years~\cite{CentralityReviews,freeman1978centrality}. 
The the most prominent ones are the connectivity degree (or just degree), closeness, betweenness, and eigenvector centrality.   
Researchers continue inventing new centrality measures and node properties from time to time to fit a particular task that was not properly covered yet~\cite{LaplacianCentrality,HybridCentrality}.   
Formulating new structural node properties, including centrality measures, for a task at hand requires domain expertise, network analysis expertise, and of course time and effort.  
It is important to devise a generic method capable of learning centrality measures provided ground truth importance of nodes.

Standard machine learning techniques can utilize previously defined centrality measures to compute a new one~\cite{grando2018approximating,mendoncca2020approximating,Zhao2020influential}. 
However, such approaches inherit the pros and cons of the centrality measures used as features and may not be able to learn an entirely new concept of centrality. 
Attempts were made to use GNN for centrality learning. 
Maurya et al.~\cite{maurya2021graph} proposed architectures that do not rely on pre-computed centrality measures but, unfortunately, they use different GNN architectures for learning closeness and betweenness measures, falling back to relying on human expertise. 
A recently proposed method utilizes the generic nature of routing betweenness centrality to learn arbitrary centrality measures~\cite{bachar2021learning}. 
Unfortunately, this approach does not scale well to large networks.

Spectral graph analysis is a widely accepted tool for graph mining, comparison, and classification~\cite{von2007tutorial,sahai2012hearing,avrachenkov2016distributed,tsitsulin2018netlsd}.
Eigenvalues represent the stationary frequencies (spectrum) of heat or wave propagation in a graph. 
The respective eigenvector components are used as a form of node representation for classification and clustering of nodes. 
In most cases this representation includes a part of the stationary state corresponding to the top $k$ eigenvalues. 
In this article we consider the full wave-form-based representation of nodes rather then their spectrum and   
use sound recognition neural network to hear their centrality.

\begin{figure}[t]
    \centering
    \includegraphics[width=\textwidth]{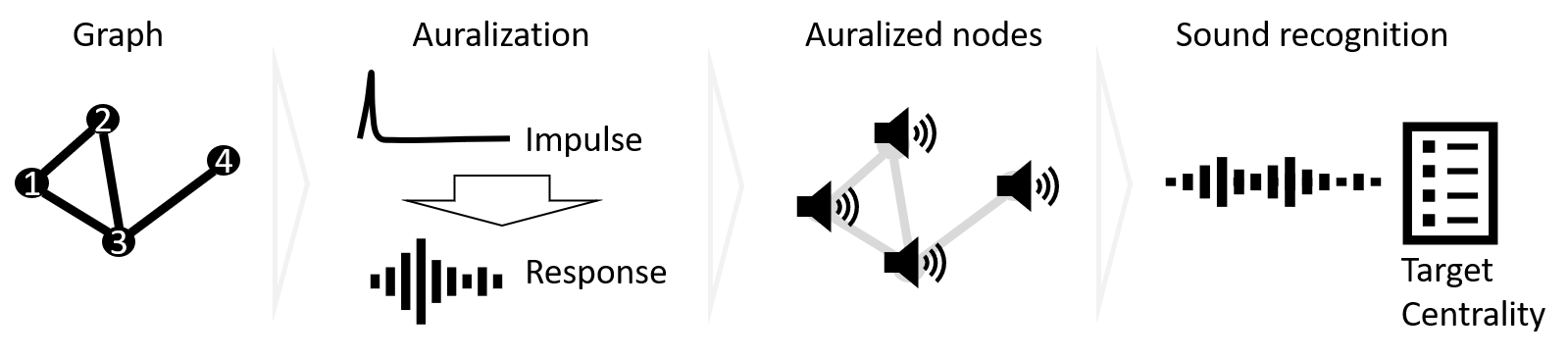}
    \caption{The pipeline of centrality learning through network auralization and sound recognition.}
    \label{fig:architecture}
\end{figure}

\paragraph{High level overview.}
The centrality learning approach proposed in this article relies on wave propagation along the links of the network.
Instead of focusing on the stationary spectrum we monitor the impulse response of the network. 
Natural reflection and interference that within the network results in complex audible wave-forms that auralize the nodes and their positions within the network. 
We use the acoustic characteristics of the nodes   
to learn centrality measures using the M5 deep neural architecture for sound recognition~\cite{dai2017very}.  
Figure~\ref{fig:architecture} summarizes the general architecture of the demonstrated centrality learning approach. 

\subsection{Summary of contributions}
Following is the list of preliminary results reported in this paper and their respective significance.
This article shows that wave-form network analysis facilitates non-trivial downstream tasks such as centrality learning. 
The results show that the M5 sound recognition neural network~\cite{dai2017very} can learn centrality measures  for variety of network models and centrality measures. 
Yet, lattice networks, in particular Watts-Strogats~\cite{watts1998collective}, are challenging as well as learning the closeness centrality measure. 
Our preliminary results advance state-of-the-art centrality learning making a few more steps toward automated inference of this important concept. 
The presented technique exemplifies the importance of  non-stationary pre-convergence state of network signal propagation. 
finally this paper demonstrates a surprising combination of graph analysis with sound recognition neural network, hoping to inspire additional architectural solutions for classification problems on graphs.  

\section{Related Work on centrality learning}
\label{sec:related_work}

\subsection{Traditional machine learning}
In recent years, with the advancement in the machine and deep learning fields, researchers utilized many of these algorithms to approximate the centrality of graph's nodes. 
Most works in this category infer some centrality measures from other centrality measures.  
In 2018, Grando et al.~\cite{grando2018approximating} rely on the degree and on the eigen vector centrality to learn other centrality measures. 
Mendonca et al.~\cite{mendoncca2020approximating} improved the work presented by Grando et al. by proposing the NCA-GE model.
The NCA-GE architecture utilizes Strucre2Vec and Graph Convolution Network (GCN) for generating a high dimensional feature vector for each node in a given graph.
To these generated node embeddings, they added the degree centrality as an additional dimension.  
Finally, they utilized the embeddings obtained to approximate node centrality. 
Zhao et al.~\cite{Zhao2020influential} detected influential nodes based on various features including  nine centrality measures.
The target measure was the influence of a node as simulated by epidemic propagation. 
Relying on pre-computed centrality measures helps inferring correlated measures, but might not generalize well to approximate un-correlated measures or tasks that require learning new measures.

\subsection{Graph neural networks}
Maurya et al.~\cite{maurya2021graph} proposed GNN-Bet and GNN-Close, two graph neural networks to approximate betweenness and closeness centralities respectively. 
Fan et al.~\cite{fan2019learning} proposed a graph neural network encoder-decoder ranking model to identify nodes having the highest betweenness. 
Last year, Bachar et al.~\cite{bachar2021learning} proposed LRC, a centrality learning architecture that relies on routing betweenness centrality~\cite{dolev2010routing}. 
While being generic LRC is heavy computation-wise and can learn centrality in graphs with only dozens of nodes.  
Moreover, it requires high quality geometric embedding to produce accurate results.

Overall centrality learning must be \emph{inductive}: trained and tested on different graphs. 
Centrality learning should be tested on real and synthetic networks with varying structures. 
It should be \emph{generic} and suitable for arbitrary target centrality measures and preferably \emph{scalable} facilitating applications on graphs larger than those it was trained on.  

\section{Methods}
\label{sec:methods}

\subsection{Network auralization}

In his subsection a simple wave-form generation process is described. 
Let $G=(V,E)$ be a simple undirected unweighted graph where $V$ is a set of $n$ nodes and $E$ is a set of $m$ edges.

Consider some quantity $s_{v,t}\in\mathbb{R}$ possessed by every node $v\in V$ at time $t$. 
Intuitively $s_{v,t}$ can be regarded as a potential of the node. 
$S_t=(s_v : v\in V)$ is the vector of potentials. 
Nodes strive to equalize their potentials by distributing energy to their neighbors. 
Please note that, although, some physical terms are used here to describe the network analysis they are not intended to discuss a real physical phenomenon and lack the rigorousity expected from a physics article.

Let $A$ denote the adjacency matrix of the graph $G$. 
Let $D=\sum_v A_v$ denote the vector of node degrees. 
We define $P_{u,v} = A_{u,v} / D_u$ as the power that $u$ may apply on $v$. 
The total power a node may apply on its neighbors is equal for all nodes.
The more neighbors a node has the less power it can apply on each one of them. 

The amount of energy every node $u$ passes to every neighbor $u$ at time $t$ is $\Delta S_{t,u,v} = S_{t-1,u}\cdot P_{u,v}$. 
In matrix form 
$\Delta S_t=Diag(S_{t-1})\times P$,
where $Diag(S_t)$ is a $n\times n$ matrix with values of $S_t$ along the main diagonal. 
The incoming energy flow of a node $v$ is $\sum_{u\in V} \Delta S_{t,u,v}$ and the outgoing energy flow is $\sum_{u\in V} \Delta S_{t,v,u}$.
Note that $\Delta S_{t,u,v}=0$ if $u$ and $v$ are not neighbors. 
Equation~\ref{eq:diffusion} describes a simple diffusion process where nodes exchange energy up to a certain fixed point, much like in the power iterations method. 
\begin{equation}
\label{eq:diffusion}
S_t = S_{t-1} + \sum_{u\in V} \Delta S_{t,u,v} - \sum_{u\in V} \Delta S_{t,v,u} 
\end{equation}
Unlike power iterations Equation~\ref{eq:diffusion} does not require normalizing the potential vector in every iteration due to energy conservation ($\sum S_{t}$ is constant). 

\begin{figure}[h]
    \centering
    \begin{subfigure}[b]{0.3\textwidth}
         \centering
         \includegraphics[width=\textwidth]{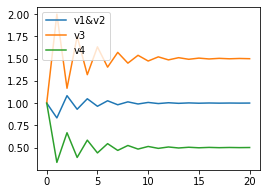}
         \caption{No momentum}
         \label{fig:oscillations-m0}
     \end{subfigure}   
    \begin{subfigure}[b]{0.3\textwidth}
         \centering
         \includegraphics[width=\textwidth]{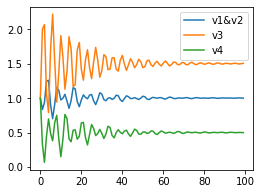}
         \caption{Momentum $m=0.9$}
         \label{fig:oscillations-m9}
    \end{subfigure}   
    \begin{subfigure}[b]{0.3\textwidth}
         \centering
         \includegraphics[width=\textwidth]{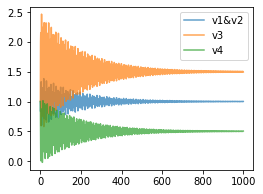}
         \caption{Momentum $m=0.99$}
         \label{fig:oscillations-m99}
     \end{subfigure}   
    \caption{Energy exchange interactions with various levels of momentum. }
    \label{fig:my_label}
\end{figure}

The stable fixed point of the energy exchange iterations is not of interest for current paper.
Let us take a close look at the dynamics of the energy exchange before the process stabilizes (see Figure~\ref{fig:oscillations-m0}).
The plot shows the potential levels of nodes from the graph in Figure~\ref{fig:architecture}.
On the first iteration the potential of $v_1$ increases the most because it has low multiple low degree neighbors. 
The potential of other nodes decreases after the first iteration because they contribute more energy than they receive. 
It is hard to see from this plot but, the energy from node $v_4$ reaches $v_1$ and $v_2$ after the second iteration and then bounces back to $v_3$ because it is the only neighbor of $v_4$. 
Nevertheless, it is clear that the location of the nodes within the network affects the magnitude and direction of the oscillations.

Next the oscillations are emphasized and the stabilization process is prolonged by retaining a portion $m$ of the energy flow from previous iteration.  
We will refer to $m$ as momentum. 
The energy flow with momentum is now represented by: 
\begin{equation}
\label{eq:uv-flow}
\Delta S_{t,u,v} = S_{t-1,u}\cdot P_{u,v} + m\cdot\Delta S_{t-1,u,v}.
\end{equation}
Note that adding momentum to $\Delta S$ does not affect the energy preservation in Equation~\ref{eq:diffusion}.

\begin{algorithm}[t]
\scriptsize
\KwIn{$m$: momentum, $A$: adjacency matrix, $l$: number of output samples}
\KwOut{$S$: the $l\times n$ matrix representing node wave forms}
\SetAlgoLined
 $\forall_{v\in V} S_{0,v}=1$ \tcp*{Impulse}
 $\forall_{u,v\in V} \Delta S_{0,u,v} = 0$\;
 $P = (A / (A.sum(dim=0)+\epsilon).T)$
 \tcp*{$\epsilon=10^{-32}$ for numeric stability.}
 \For{$t \in [1,\ldots,l]$}{ 
    $\Delta S_{t} = Diag(S_{t-1}) \times P + m\cdot \Delta S_{t-1}$\; 
    $S_t = S_{t-1} + \Delta S_{t}.sum(dim=0) - \Delta S_{t}.sum(dim=1)$\ \tcp*[r]{Response}
 }
 $S = (S.T - S.T.mean(dim=0)).T$ \tcp*{Remove the DC component of the wave-forms.}
 \KwRet{S}
 \caption{Network auralization}
\label{alg:auralization}
\end{algorithm}

Figures~\ref{fig:oscillations-m9} and~\ref{fig:oscillations-m99} show the oscillations with $m=0.9$ and $m=0.99$ respectively. 
We can see that the stabilization process is significantly prolonged. 
We can also see in Figure~\ref{fig:oscillations-m9} irregularities caused by interference and reflection as explained in the spectral analysis literature.   
Setting $m=1$ will prevent the process from stabilizing. 

Algorithm~\ref{alg:auralization} presents the pseudo code of network auralization adapted for PyTorch implementation.%
\footnote{The full source code is available on GitHub: \url{https://github.com/puzis/centrality-learning}}
The operator $T$ is matrix transpose. 
The operators $sum$ and $mean$ aggregate elements of a matrix along the dimension specified by $dim$. 
The DC component of the output wave-forms (values on which $S$ stabilizes after impulse response) is not of interest. 
It may also hinder the convergence of sound recognition models.
Furthermore, it is a centrality measure on its own, roughly corresponding to eigenvector centrality. 
Thus we remove the DC component in Line~8 of Algorithm~\ref{alg:auralization} in order show that the wave-form itself bears significant information about the location of a node within the graph.

\subsection{Centrality learning as a sound recognition problem}
\subsubsection{Architecture}
Sound is the most common and most studied form of waves. 
Many deep convolutional neural networks were developed in the past years to recognize speech~\cite{bahar2019comparative}, emotions~\cite{abdullah2021multimodal}, background sounds~\cite{dai2017very}, etc. 
Current study relies on the M5 very deep convolutional neural network proposed by Dai et al. for recognition of environmental sounds in urban areas~\cite{dai2017very}. 
There are two important details about their neural network architecture that should be mentioned here.
First, the filter size of their first convolutional layer is set to 80, a sufficiently large value to cover the common wavelengths of natural sounds. Second, their last layers include global pooling and softmax activation function with 10 outputs to produce a classifier with 10 target classes.

In current study the M5 classifier is transformed into a regression architecture by replacing the softmax with a fully connected linear layer.
Preliminary experiments with various activation functions showed that linear activation produces the best results in terms of Pearson correlation coefficient.  
Also not tested with other sound recognition architectures replacing softmax with a fully connected linear layer is a common tweak for turning a classifier into a regression model. 
We consider the M5 regressor as a function that maps node's wave-form to a real number: $M5: \mathbb{R}^{l} \rightarrow \mathbb{R}$. 
The learned centrality measure of the node $v\in V$ is therefore $M5(S_{\bullet,v})$ where $\bullet$ represents all possible values.  
$M5(S)$ is the centrality vector of all nodes in $G$.

\subsubsection{Objective function}

Let $C$ denote the target centrality measure and $P$ denote the predicted centrality values.  
The target variable for training the M5 regressor was chosen to be the Pearson correlation coefficient: 
$$
\rho(C, P) = \frac{cov(C,P)}{std(C)\cdot std(P)} 
$$. 
The loss for training the M5 neural network is:
\begin{equation}
    \label{eq:loss}
    loss = 1 - \rho(C, M5(S))
\end{equation}

Correlations are common performance indicators for centrality learning~\cite{grando2018approximating,mendoncca2020approximating,maurya2021graph,bachar2021learning}. 
Pearson correlation coefficient has the most accessible differentiable implementation in the PyTorch deep learning library. 
Other correlation coefficients can be used as long as they have differentiable implementations.   
A notable one, is differential implementation of Spearman correlation coefficient~\cite{blondel2020fast}.

\subsubsection{The training procedure}
In this subsection we consider the general task of centrality learning. 
Connectivity degree (Deg), closeness centrality (CC), eigenvector centrality (EC), and betweenness centrality (CC) are considered as the target centrality measures to demonstrate the learning process.

\begin{figure}[t]
    \centering
    \begin{subfigure}[b]{0.19\textwidth}
         \centering
         \includegraphics[width=\textwidth]{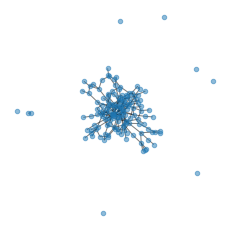}
         \caption{}
         \label{fig:gnm}
    \end{subfigure}   
    \begin{subfigure}[b]{0.19\textwidth}
         \centering
         \includegraphics[width=\textwidth]{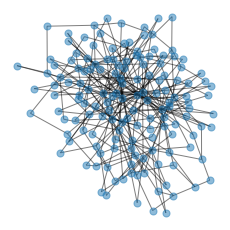}
         \caption{}
         \label{fig:ba}
    \end{subfigure}   
    \begin{subfigure}[b]{0.19\textwidth}
         \centering
         \includegraphics[width=\textwidth]{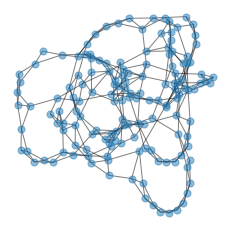}
         \caption{}
         \label{fig:}
    \end{subfigure}   
    \begin{subfigure}[b]{0.19\textwidth}
         \centering
         \includegraphics[width=\textwidth]{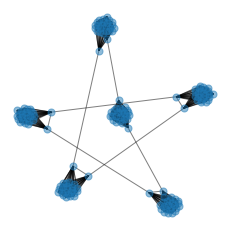}
         \caption{}
         \label{fig:caveman}
    \end{subfigure}   
    \begin{subfigure}[b]{0.19\textwidth}
         \centering
         \includegraphics[width=\textwidth]{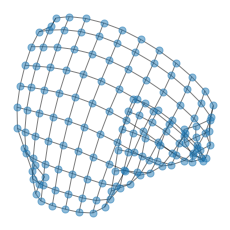}
         \caption{}
         \label{fig:grid}
    \end{subfigure}  
    \\
    \begin{subfigure}[b]{0.19\textwidth}
         \centering
         \includegraphics[width=\textwidth]{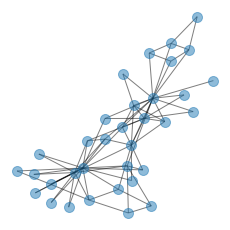}
         \caption{}
         \label{fig:karate}
    \end{subfigure}   
    \begin{subfigure}[b]{0.19\textwidth}
         \centering
         \includegraphics[width=\textwidth]{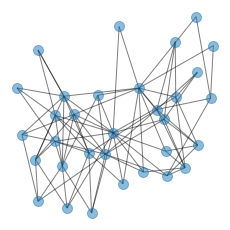}
         \caption{}
         \label{fig:davis}
    \end{subfigure}   
    \begin{subfigure}[b]{0.19\textwidth}
         \centering
         \includegraphics[width=\textwidth]{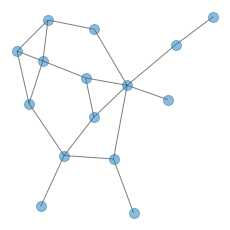}
         \caption{}
         \label{fig:florentine}
    \end{subfigure}   
    \begin{subfigure}[b]{0.19\textwidth}
         \centering
         \includegraphics[width=\textwidth]{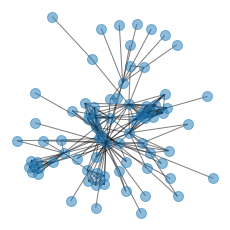}
         \caption{}
         \label{fig:miserables}
    \end{subfigure}   
    \begin{subfigure}[b]{0.19\textwidth}
         \centering
         \includegraphics[width=\textwidth]{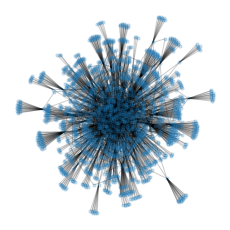}
         \caption{}
         \label{fig:internet}
    \end{subfigure}  
    \caption{Examples of some random and well known networks. On the top: Erdos-Renyi (ER) random graph~\cite{erdHos1960evolution}, Barabasi-Albert (BA) scale-free graph~\cite{barabasi1999emergence}, Watts-Strogatz (WS) small-world graph~\cite{watts1998collective}, connected caveman graph~\cite{watts1999networks}, and a regular grid. On the bottom: Karate club~\cite{zachary1977information}, Southern women graph~\cite{davis2009deep}, Florentine families graph~\cite{breiger1986cumulated}, Les Miserables~\cite{knuth1993stanford}, Autonomous Systems level random Internet graph~\cite{elmokashfi2010scalability} } 
    \label{fig:graph-examples}
\end{figure}

It is evident from past research and from the results in this article that certain kinds of networks are harder to learn than others. 
Therefore the models are trained on five different random network models
shown in Figure~\ref{fig:graph-examples} (a-e).
All training was performed on small graphs of 150 nodes and various densities.

The training phase was split to epochs (see Algorithm~\ref{alg:training}). 
In every epoch a set of small random graphs were generated.   
The number of graphs grows linearly with epochs (line~3). 
Graphs are generated from all the five models considered for training. 
Graphs are auralized in line~8.   
Next, ten iterations are made to optimize the parameters of the M5 regression model. 
The numbers in lines 3 and 9 were chosen empirically and are subject to experiments in future research. 
Multiple optimization steps on a set of random graphs (lines 9-12) are required, otherwise, the M5 model parameters fail to converge. 
As the model converges the batch size (number of random graphs) need to increase in order to reduce the noise in the optimization process.

\begin{algorithm}[t]
\scriptsize
\SetAlgoLined
 \For{$epoch \in [1,\ldots,300]$}{ 
    $Gs\leftarrow\emptyset$\tcp*[l]{graphs for current epoch}
    \For{$10+\lfloor\frac{epoch}{10}\rfloor$ times}{ 
        $Gen\leftarrow$ pick a random graph generator\; 
        $G\leftarrow$ generate a random graph using $Gen$\;
        $Gs\leftarrow Gs\cup\{G\}$\;
    } 
    \lFor{$G\in Gs$}{$S(G)\leftarrow$ network auralization ($G$)}
    \For(\tcp*[h]{10 batches on the same graphs}){10 times}{
        $batchloss\leftarrow \frac{1}{10\cdot|Gs|}\sum_{G\in Gs} (1-\rho(M5(S(G))))$\;        $batchloss$ backprogation\;
        optimizer step\;
    }
 }
 \KwRet{S}
 \caption{The centrality learning process}
\label{alg:training}
\end{algorithm}

\subsubsection{The testing procedure.}
New random graphs are generated for the testing phase.
Three types of test sets are considered: 
\\\noindent 1. Small random graphs ($n=150$) similar to the training process. 
\\\noindent 2. Larger random graphs ($n=1500$) generated using the same models. 
\\\noindent 3.  Four famous graphs and a random Internet topology as depicted in Figure~\ref{fig:graph-examples} (f-j). \\\noindent 
Pearson correlation was computed between the centrality leaned using the M5 sound recognition neural network and the ground truth centrality measures for each one of the test graphs.

Models trained on the small networks were applied on the large networks as well. 
This is possible due to the size-invariant representation of the nodes' wave-forms in time domain.  
The wave-forms of larger networks usually exhibit more frequencies (a richer sound) corresponding to a larger number of eigenvalues. 
We use fixed-size time series of 10K samples as the input for the M5 sound recognition network. 

\section{Results and discussion}

\subsection{The voice of graphs and nodes}
\label{sec:voice}

\begin{figure}[t]
    \centering
    \begin{subfigure}[b]{1\textwidth}
         \centering
         \includegraphics[width=0.7\textwidth]{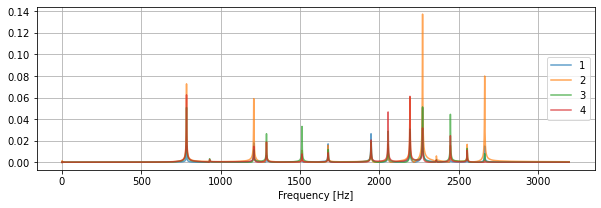}
         \includegraphics[width=0.28\textwidth]{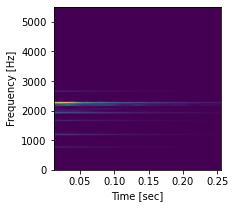}
        \caption{Florentine families \includemedia[
                addresource=soundflorentine.wav,
                flashvars={
                source=soundflorentine.wav
                &autoPlay=true
                },
                transparent,
                passcontext 
                ]{\color{blue}\framebox[5.5cm][c]{Click to play (all nodes, 13 seconds)}}{APlayer.swf}}
    \end{subfigure}   
    \begin{subfigure}[b]{1\textwidth}
         \centering
         \includegraphics[width=0.7\textwidth]{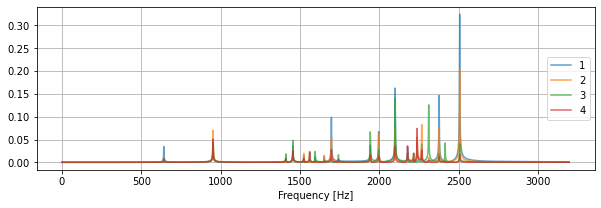}
         \includegraphics[width=0.28\textwidth]{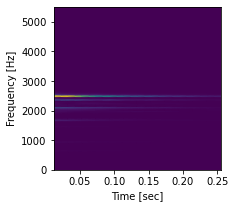}
        \caption{Karate club \includemedia[
                addresource=soundkarate.wav,
                flashvars={
                source=soundkarate.wav
                &autoPlay=true
                },
                transparent,
                passcontext 
                ]{\color{blue}\framebox[5.5cm][c]{Click to play (all nodes, 30 seconds)}}{APlayer.swf}}
    \end{subfigure}      
    \begin{subfigure}[b]{1\textwidth}
         \centering
         \includegraphics[width=0.7\textwidth]{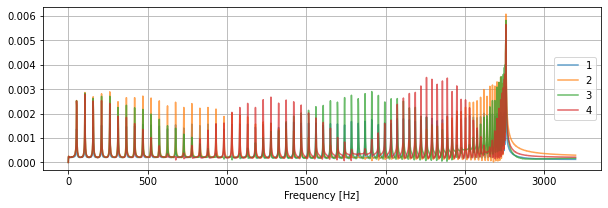}
         \includegraphics[width=0.28\textwidth]{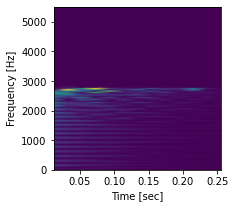}
        \caption{Line graph with 150 nodes \includemedia[
                addresource=soundline.wav,
                flashvars={
                source=soundline.wav
                &autoPlay=true
                },
                transparent,
                passcontext 
                ]{\color{blue}\framebox[5.5cm][c]{Click to play (20 nodes, 18 seconds)}}{APlayer.swf}}
    \end{subfigure}      
    \caption{The voice of graphs and nodes. Left: the spectra of four nodes in a graph. Right: a spectrogram of one arbitrary node.}
    \label{fig:sound}
\end{figure}

The echo chambers formed by graphs produce sound patterns as can be seen and heard in the multimedia Figure~\ref{fig:sound}. 
On the left side of the spectrum of nodes depicted. 
While peak frequencies remains the same for different nodes in the graph their intensities vary.   
The proposed centrality learning approach relies on these differences. 

\subsection{Centrality learning results}


\begin{figure}[t]
    \centering
    \includegraphics[width=0.49\textwidth]{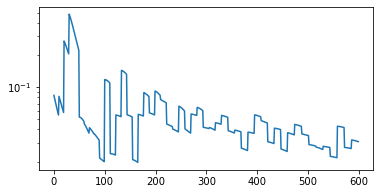}
    \includegraphics[width=0.49\textwidth]{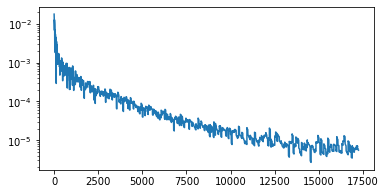}
    \caption{The loss ($1-\rho(M5(S))$) as a function of the number of optimization steps when learning the connectivity degree. Left: 50 epochs. Right: 500 epochs.}
    \label{fig:loss}
\end{figure}

Figure~\ref{fig:loss} shows the training loss while learning nodes' connectivity degree from their auralization.   
It is clear that node auralization contains information about connectivity degree.

Table~\ref{tab:pearson} presents the Pearson correlation coefficients while testing the trained M5 sound recognition model.  
Naturally, degree is the easiest to learn, but the correlation coefficients for other centrality measures are generally better than the respective correlations with other centrality measures (0.48-0.91 according to \cite{bachar2021learning}).   
This fact suggests that the wave-forms produced by Algorithm~\ref{alg:auralization} encode information about the position of nodes within their networks.

Grid and Cavemen graphs show extreme results due to an over-fitting and sample leakage from train to test due to their low variability. 
The M5 network memorizes the node wave-forms and fails to extrapolate the learned centrality to larger networks. 
This result highlights the fact that the model successfully extrapolates to networks an order of magnitude larger than those it was trained on. 

While performing reasonably well on most small real world networks and on the random Internet graph, the learned model fails on the Les Miserables social network. 
Even the predicted degree of nodes is loosely correlated with the actual degree. 
I could not find a reasonable explanation for such  behavior. 

\begin{table}[t]
    \centering
    \includegraphics[width=0.9\textwidth]{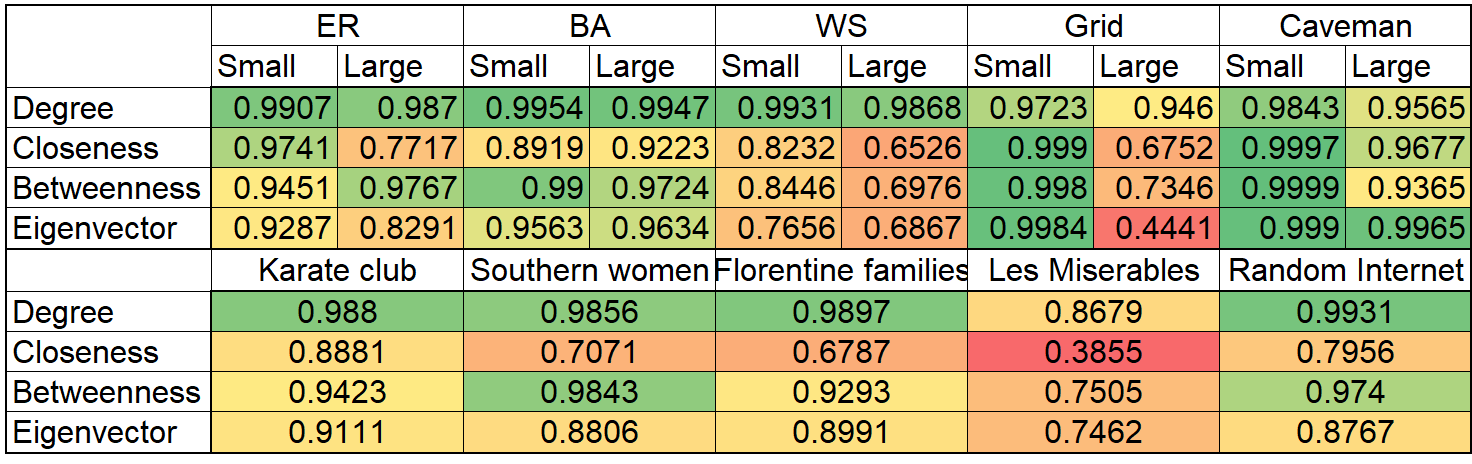}
    \caption{Performance of the M5 sound recognition neural network when predicting centrality of auralized nodes on various network. }
    \label{tab:pearson}
\end{table}

\section{Conclusion}
\label{sec:conclusion}
In this paper, demonstrates a surprising application of sound recognition neural network for learning centrality measures from auralized nodes in graphs.     
The demonstrated model learns to infer centrality from the sound of nodes.
The model extrapolates well to networks larger than network in its training set.     
This article also demonstrates by an example that the stabilization process of graph algorithms bears information about the graph structure.  
It may form a basis for new types of message passing graph neural networks where the energy exchange parameters are learned during the training process. 
Future work includes investigating a transfer learning problem where centrality learned from one kind of graphs (e.g. ER) is adjusted to fit graphs with significantly different structure (e.g. WS). 
Another theme with a lot of fun and unexplored research opportunities is natural graph auralization, not to be confused with artificial/engineered sonification. 

\bibliographystyle{unsrt}
\bibliography{references}

\end{document}